\documentclass[twocolumn]{aa}
\usepackage{graphicx}
\usepackage{natbib}
\def\pmb#1{\setbox0=\hbox{#1}%
 \kern-.025em\copy0\kern-\wd0
 \kern.05em\copy0\kern-\wd0l
 \kern-.025em\raise.0433em\box0}
\def\eg{{\it e.g.\ }}

\def\ie{{\it i.e.\ }}

\def\del{\delta}                
\def\hpc{$h^{-1}$Mpc }
\def\hpcp{$ h^{-1}$Mpc. }
\def\hpcv{$ h^{-1}$Mpc, }

\def\delo{\delta^{O}}
\def\delp{\delta^{P}}
\def\delr{\delta^{R}}

\def\z{$\,${\it z}$\,$}

\def\eg{{e.g.,~}}
\def\ie{{i.e.~}}
\def\lsim{\raise0.3ex\hbox{$<$}\kern-0.75em{\lower0.65ex\hbox{$\sim$}}} 
\def\gsim{\raise0.3ex\hbox{$>$}\kern-0.75em{\lower0.65ex\hbox{$\sim$}}} 
\def\lesssim{\mathrel{\hbox{\rlap{\hbox{\lower4pt\hbox{$\sim$}}}\hbox{$<$}}}}
\def\gtrsim{\mathrel{\hbox{\rlap{\hbox{\lower4pt\hbox{$\sim$}}}\hbox{$>$}}}}
\def\sc1{\raise2.1ex\hbox{\tiny $r\!\!=\!\!4$}\kern-0.95em{\hbox{$=$}}}

\newcommand{\unit}[1]{\ifmmode \:\mbox{\rm #1}\else \mbox{#1}\fi}

\def\ltsima{$\; \buildrel < \over \sim \;$}
\def\simlt{\lower.5ex\hbox{\ltsima}}
\def\gtsima{$\; \buildrel > \over \sim \;$}
\def\simgt{\lower.5ex\hbox{\gtsima}}          

\def\sc{{\rm Science\ }}

\begin{document}

\title{The VIMOS VLT Deep Survey\thanks{Based on observations obtained
         with the European Southern Observatory Very Large Telescope,
         Paranal, Chile. This paper also uses observations obtained
         with MegaPrime/MegaCam, a joint project of CFHT and
         CEA/DAPNIA, at the Canada-France-Hawaii Telescope (CFHT)
         which is operated by the National Research Council (NRC) of
         Canada, the Institut National des Science de l'Univers of the
         Centre National de la Recherche Scientifique (CNRS) of
         France, and the University of Hawaii. This work is based in
         part on data products produced at TERAPIX and the Canadian
         Astronomy Data Centre as part of the Canada-France-Hawaii
         Telescope Legacy Survey, a collaborative project of NRC and
         CNRS.}: The build-up of the colour-density relation }

\author{
O. Cucciati \inst{1,2}
\and A. Iovino \inst{1}
\and C. Marinoni \inst{1,3}
\and O. Ilbert \inst{4,5}
\and S. Bardelli  \inst{6}
\and P. Franzetti \inst{7}
\and O. Le F\`evre \inst{5}
\and A. Pollo \inst{5}
\and G. Zamorani \inst{6} 
\and A. Cappi    \inst{6}
\and L. Guzzo \inst{1}
\and H.J. McCracken \inst{8,9}
\and B. Meneux \inst{7,1}
\and R. Scaramella \inst{10,11}
\and M. Scodeggio \inst{7}
\and L. Tresse \inst{5}
\and E. Zucca    \inst{6}
\and D. Bottini \inst{7}
\and B. Garilli \inst{7}
\and V. Le Brun \inst{5}
\and D. Maccagni \inst{7}
\and J.P. Picat \inst{12}
\and G. Vettolani \inst{10}
\and A. Zanichelli \inst{10}
\and C. Adami \inst{5}
\and M. Arnaboldi \inst{13}
\and S. Arnouts \inst{5}
\and M. Bolzonella  \inst{6} 
\and S. Charlot \inst{14,8}
\and P. Ciliegi    \inst{6}  
\and T. Contini \inst{12}
\and S. Foucaud \inst{7}
\and I. Gavignaud \inst{12,15}
\and B. Marano     \inst{4}  
\and A. Mazure \inst{5}
\and R. Merighi   \inst{6} 
\and S. Paltani \inst{16,17}
\and R. Pell\`o \inst{12}
\and L. Pozzetti    \inst{6} 
\and M. Radovich \inst{13}
\and M. Bondi \inst{10}
\and A. Bongiorno \inst{4}
\and G. Busarello \inst{13}
\and S. de la Torre \inst{5}
\and L. Gregorini \inst{10}
\and F. Lamareille \inst{12}
\and G. Mathez \inst{12}
\and Y. Mellier \inst{8,9}
\and P. Merluzzi \inst{13}
\and V. Ripepi \inst{13}
\and D. Rizzo \inst{12}
\and S. Temporin \inst{1}
\and D. Vergani \inst{7}
}
   
\offprints{O. Cucciati, e-mail: olga.cucciati@brera.inaf.it}
 
\institute{
INAF-Osservatorio Astronomico di Brera - Via Brera 28, Milan,
Italy
\and
Universit\'a di Milano-Bicocca, Dipartimento di Fisica - 
Piazza della Scienza, 3, I-20126 Milano, Italy
\and
Centre de Physique Th\'eorique, UMR 6207 CNRS-Universit\'e de Provence, 
F-13288 Marseille, France 
\and
Universit\`a di Bologna, Dipartimento di Astronomia - Via Ranzani,1,
I-40127, Bologna, Italy
\and
Laboratoire d'Astropysique de Marseille, UMR 6110 CNRS-Universit\'e de
Provence,  BP8, 13376 Marseille Cedex 12, France
\and
INAF-Osservatorio Astronomico di Bologna - Via Ranzani,1, I-40127, Bologna, Italy
\and
IASF-INAF - via Bassini 15, I-20133, Milano, Italy
\and
Institut d'Astrophysique de Paris, UMR 7095, 98 bis Bvd Arago, 75014
Paris, France
\and
Observatoire de Paris, LERMA, 61 Avenue de l'Observatoire, 75014 Paris, 
France
\and
IRA-INAF - Via Gobetti,101, I-40129, Bologna, Italy
\and
INAF-Osservatorio Astronomico di Roma - Via di Frascati 33,
I-00040, Monte Porzio Catone,
Italy
\and
Laboratoire d'Astrophysique de l'Observatoire Midi-Pyr\'en\'ees (UMR 
5572) -
14, avenue E. Belin, F31400 Toulouse, France
\and
INAF-Osservatorio Astronomico di Capodimonte - Via Moiariello 16, I-80131, Napoli,
Italy
\and
Max Planck Institut fur Astrophysik, 85741, Garching, Germany
\and
European Southern Observatory, Karl-Schwarzschild-Strasse 2, D-85748
Garching bei M\"unchen, Germany
\and
Integral Science Data Centre, ch. d'\'Ecogia 16, CH-1290 Versoix, Switzerland
\and
Geneva Observatory, ch. des Maillettes 51, CH-1290 Sauverny, Switzerland
}

\authorrunning{Cucciati et al.}
\titlerunning{The VVDS: The build-up of the colour-density relation}

\date{Received: March 8, 2006, Accepted: July 19, 2006}

\abstract{ We investigate the redshift and luminosity evolution of the
galaxy colour-density relation using the data from the First Epoch
VIMOS-VLT Deep Survey (VVDS). The size (6582 galaxies with good
quality redshifts), depth ($I_{AB}\leq 24$) and redshift sampling rate
(20\% on the mean) of the survey enable us to reconstruct the 3D
galaxy environment on relatively local scales ($R=5$\hpc) up to
redshift $\z \sim 1.5$.  Particular attention has been devoted to
calibrate a density reconstruction scheme, which factors out survey
selection effects and reproduces in an unbiased way the underlying
``real'' galaxy environment.  We find that the colour-density relation
shows a dramatic change as a function of cosmic time. While at lower
redshift we confirm the existence of a steep colour-density relation,
with the fraction of the reddest(/bluest) galaxies of the same
luminosity increasing(/decreasing) as a function of density, this
trend progressively disappears in the highest redshift bins
investigated.  Our results suggest the existence of an epoch
(more remote for brighter galaxies) characterized by the
absence of the colour-density relation on the $R=5$\hpc scales
investigated. The rest frame $u^{*}-g'$ colour-magnitude diagram shows
a bimodal pattern in both low and high density environments up to
redshift $z\sim 1.5$. We find that the bimodal distribution is not
universal but strongly depends upon environment: at lower redshifts
the colour-magnitude diagrams in low and high density regions are
significantly different while the progressive weakening of the
colour-density relation causes the two bimodal distributions to nearly
mirror each other in the highest redshift bin investigated.  Both the
colour-density and the colour-magnitude-density relations, on the 
$R=5$\hpc scales, appear to be a transient, cumulative product of 
genetic and environmental factors that have been operating over at 
least a period of 9 Gyr.  These findings support an evolutionary 
scenario in which star formation/gas depletion processes are 
accelerated in more luminous objects and in high density environments: 
star formation activity is progressively shifting with cosmic time 
towards lower luminosity galaxies (downsizing), and out of high density
environments.
   
}
  
\maketitle

\section{Introduction}\label{intro}

There is a well known connection between galaxy properties such as
morphology, luminosity, integral colour, specific star formation rate
(SFR), surface brightness and the local environment wherein galaxies
reside
\citep[e.g.,][]{spiz,dressler1980,withmore1993,poggianti1999,marlum,
marhu,balogh2004a,balogh2004b,hogg,blanton2005,weinmann06}.  These
correlations extend smoothly over a wide range of density
enhancements, from the extreme environment of rich clusters to very
low densities, well beyond the region where the cluster environment is
expected to have much influence \citep[e.g.,
][]{postman1984,zabludoff1998,gomez}.

Despite increasing precision in quantifying environmental
correlations, we still lack a satisfactory understanding of where
these trends stem from.  Several physical mechanisms are expected to
be crucial in determining the properties of galaxies in over-dense
regions: ram pressure stripping of gas \citep{gunn_gott1972},
galaxy-galaxy merging \citep{toomre1972}, strangulation
\citep{larson1980} and harassment \citep{moore1996}. Although these
processes are plausible, each mechanism has specific environments and
timescales in which it operates most effectively, and additional
ingredients may turn out to be essential for progressing towards a
coherent physical interpretation of the observations.  For example,
environmental analyses have yet to elucidate the
relative and complementary importance of the physics regulating galaxy
formation. It is not yet clear to what extent local phenomena such as
feedbacks from supernovae and central black holes do contribute to the
observed density dependence of the galaxy structural parameters.
Moreover, it is known that in a Gaussian random field there is a
statistical correlation between mass fluctuations on different scales,
with most massive halos preferentially residing within over-densities
on larger scales \citep[see][]{kaiser1987, mo1996}.  What is less
clear is the role of initial cosmological conditions in triggering the
observed density dependence of optical galaxy properties
\citep{abbas2005}.

A key question that still needs to be addressed is whether these
environmental dependencies were established early on when galaxies
first assembled, or whether they are the present day cumulative end
product of multiple processes operating over a Hubble time. In other
words if these dependencies arise during the formation of galaxies
(the so-called `nature' hypothesis) or whether they are caused by
density-driven evolution (the `nurture' scenario).

A promising approach to addressing these issues involves extending
observations beyond the local universe. The relations between
environment and galaxy properties have still virtually no empirical
constraints beyond $z\sim0.5$, except in cluster of galaxies
\citep[e.g.,][]{smith05,postman05,tanaka}.  A preliminary analysis
over volumes which average over many different environments has been
attempted out to $z=1$ by \citet{nuijten2005}, using a photometric
sample of galaxies. However it has been recently stressed that
environmental investigations crucially require high resolution
spectral measurements of galaxy positions.  For example,
\citet{cooper2005} find that even optimistic photometric redshift
errors ($\sigma=0.02$) smear out the galaxy distribution irretrievably
on small scales, significantly limiting the application of photometric
redshift surveys to environment studies \citep[but see for a
particular case][]{guzzo2006}.

Large and deep redshift surveys of the universe are the best available
instrument to select a representative sample of the galaxy population
over a broad and continuous range of densities and cosmic
epochs. Moreover these surveys open up the possibility of exploring
such trends in different magnitude bands.
In this study, we use the VIMOS VLT Deep Survey \citep{lefevre2005a}, 
the largest ($6582$ objects), deepest 
($0.25<z<1.5$), purely-magnitude selected ($I_{AB}\leq 24$)
redshift sample currently available, to explore the colour-density
relation as a function of both luminosity and cosmic time.
In particular the main goal of this investigation 
is to portray the colour-density relation 
at different epochs and evaluate eventual changes in its overall
{\it normalization} \citep[Butcher \& Oemler effect,][]{beo} and 
{\it slope} \citep[Dressler effect,][]{spiz,dressler1980}.
 
While redshift surveys have grown in scale and environmental studies
have acquired momentum, much less attention has been devoted to
investigate how the various systematics introduced by the particular
survey observing strategies may affect the estimation of
environment. In our study, we pay special attention to constrain the
parameter space where the VVDS density field
reproduces in a statistically unbiased way the underlying parent
density field.

This paper is set out as follows: in \S \ref{sample} we briefly
describe the first-epoch VVDS-0226-04 data sample. In \S \ref{method}
we introduce the technique applied for reconstructing the
three-dimensional density field traced by VVDS galaxies, providing
details about corrections for various selection effects. In \S
\ref{simul} we test the statistical representativity of the reconstructed 
VVDS density field using mock catalogues.  We present our results on the
dependence of galaxy colours from local density in \S \ref{results} and
discuss them in \S \ref{discussion}.  Conclusions are drawn in \S
\ref{conclusion}.

The coherent cosmological picture emerging from independent
observations and analyses motivates us to frame all the results
presented in this paper in the context of a flat, vacuum dominated
cosmology with $\Omega_m=0.3$ and $\Omega_{\Lambda}=0.7$.  Throughout,
the Hubble constant is parameterized via $h=H_{0}/100$.  All
magnitudes in this paper are in the AB system \citep{oke_gunn83}, and
from now on we will drop the suffix AB.

\section{The First-Epoch VVDS  Redshift Sample}\label{sample} 

The primary observational goal of the VIMOS-VLT Deep Survey as
well as the survey strategy and first-epoch observations in the
VVDS-0226-04 field (from now on simply VVDS-02h) are presented in
\citet{lefevre2005a}, hereafter Paper I.

Here it is enough to stress that, in order to minimize selection
biases, the VVDS survey in the VVDS-02h field has been conceived as a
purely flux-limited ($17.5\leq I \leq24$) survey, i.e. no target
pre-selection according to colours or compactness was implemented.
Stars and QSOs have been {\it a-posteriori} removed from the final
redshift sample. Photometric data in this field are complete and free
from surface brightness selection effects, up to the limiting
magnitude $I=24$ \citep{mccracken2003,lefevre2004b}. $B,V,R,I$
photometry was acquired with the wide-field 12K mosaic camera at the
CFHT, while $u^{*},g',r',i',z'$ photometry is part of the
Canada-France-Hawaii Telescope Legacy Survey.

First-epoch spectroscopic observations in the VVDS-02h field were
carried out using the VIMOS multi-object spectrograph
\citep{lefevre2003} during two runs between October and December 2002
(see Paper I).  VIMOS observations have been performed using 1
arcsecond wide slits and the LRRed grism, which covers the spectral
range $5500<\lambda(\AA)<9400$ with an effective spectral resolution
$R\sim 227$ at $\lambda=7500\AA$.  The accuracy in redshift
measurements is $\sim$275 km/s.  Details on observations and data
reduction are given in Paper I, and in \citet{lefevre2004}.

The first-epoch VVDS-02h data sample extends over a sky area of
0.7$\times$0.7 $deg^2$, which was targeted according to a 1, 2 or 4 passes
strategy, \ie giving to any single galaxy in the field 1, 2 or 4
chances to be targeted by VIMOS (see Figure 12 of paper I), and has a
median redshift of about \z$\sim$0.76.  It contains 6582 galaxies with
secure redshifts, \ie redshift determined with a quality flag$\ge$2,
see Paper I (5882 with $0.25 \leq z \leq 1.5$) and probes a comoving
volume (up to \z=1.5) of nearly $1.5\cdot 10^6 h^{-3}$Mpc$^{3}$ in a
standard $\Lambda$CDM cosmology. This volume has transversal
dimensions $\sim$ 37$\times$37 \hpc at \z=1.5 and extends over 3060
\hpc in radial direction.

For this study we define also a sub-sample (VVDS-02h-4P) with galaxies
selected in a contiguous sky region of
0.4$\times$0.4 $deg^2$ which has been homogeneously targeted four times by
VIMOS observations.  The VVDS-02h-4P subsample contains 2903 galaxies
with secure redshift (2647 with $0.25 \leq z \leq 1.5$) and probes
one-third of the total VVDS-02h volume.

\section{Environment Reconstruction Scheme}\label{method} 

To study environmental effects on galaxy properties, we need to define
an appropriate density estimator which properly corrects for 
all the survey selection biases.

We characterize the environment surrounding a given galaxy at comoving
position {\bf r}, by means of the dimensionless 3D density contrast
smoothed over a typical dimension $R$:

\begin{equation} \displaystyle \del({\bf r},R) =
\frac{\rho({\bf r},R)-\overline{\rho}({\bf r})}{\overline{\rho}({\bf r})},
\label{defdg} \end{equation}

\noindent where $\rho({\bf r},R)$ is the number density of galaxies
brighter than a fixed absolute magnitude threshold $\mathcal{M}^c$ and
$\overline{\rho}({\bf r})$ is the mean density at position {\bf
r}. The smoothed number density of galaxies on a scale R is estimated
as an appropriately weighted convolution between Dirac's delta
functions and some arbitrary filter F:

\begin{equation} \displaystyle \rho({\bf
r},R,<\mathcal{M}^c)=\sum_i \frac{ \del^{D}({\bf r}-{\bf
r_i})*F\big(\frac{|{\bf
r-r_i}|}{R}\big)}{S(r_i,\mathcal{M}^c)\Phi(m)\zeta(z,m)\Psi(\alpha,\delta)}, 
\label{defrg} \end{equation}

\noindent where the sum is taken over all the galaxies excluding that at
position {\bf r} (the one for which we estimate the surrounding
environment) and where $F(|{\bf r}|/R)$ is the window-function, which
in this study is modeled in terms of a normalized Gaussian filter:

\begin{equation} \displaystyle F\Big(\frac{|\Delta {\bf r}|}{R}\Big)= \frac{1}{(2 \pi
R^{2})^{3/2}} exp\left[-\frac{1}{2}\left(\frac{|\Delta {\bf r}|}{R}\right)^2\right].  \label
{wg}
\end{equation}

Note that, for example, with the choice R=5\hpc we smooth the galaxy
distribution over an effective volume which roughly corresponds to
that enclosed in spheres of radius 8\hpc.

As shown by \citet{ilbert2005}, $\overline{\rho}({\bf r})$ evolves by
nearly a factor of 2 for galaxies brighter than $\mathcal{M}^*_B(z=0)$
from redshift $z=0$ to redshift $z=1$.  We thus compute the
characteristic mean density at position ${\bf r}$ with equation \ref{defrg} by 
simply averaging the galaxy distribution in survey slices $r\pm R_s$, 
with $R_s=400$ \hpc \citep[see][]{marinoni2005}.

Finally, the four functions in the denominator of equation \ref{defrg}
correct for various survey observational characteristics:

\begin{itemize}

\item[-] $S(r, \mathcal{M}^c)$ is the distance-dependent selection
function of the sample which corrects for the sample progressive
radial incompleteness. Since our spectroscopic sample is limited at
bright and faint apparent magnitudes (17.5$\leq$I$\leq$24), at any
given redshift we can only observe galaxies in a specific,
redshift-dependent, absolute magnitude range. The function $S(r,
\mathcal{M}^c)$ is computed using the galaxy luminosity function
derived by \citet{ilbert2005} and assuming $\mathcal{M}^c = -15+5 \log h$. 
A thorough discussion of this function can be found in
\citet{marinoni2005}.  

\item[-]  $\Phi(m)$  corrects for the slight bias against bright
objects introduced by the slit positioning tool VMMPS/SPOC
\citep{bottini2005}.
 
\item[-] $\zeta(z,m)$ is the correction for the varying
spectroscopic success rate as a function of the apparent $I$
magnitude and of z itself \citep[see][]{ilbert2005}.

\item[-] $\Psi(\alpha,\delta)$ is a correction for the uneven
spectroscopic sampling of the VVDS on the sky, depending on the
different number of passes done by the VIMOS spectrograph in different
sky regions.

\end{itemize}

The density field reconstructed using eq. \ref{defrg} has the
advantage of exploiting all the galaxies in our sample. In principle,
the environment is thus defined as the fluctuation field traced at
every redshift by galaxies as fainter as $\mathcal{M}^c = -15+5 \log h$.

The assumption implicit in this reconstruction scheme is that the
subset of galaxies luminous enough to enter our flux-limited sample at
a given redshift are fair tracers of the full population of
galaxies. With this assumption we neglect possible biases due to the 
dependence of clustering on luminosity; moreover, adopting a universal 
luminosity function we do not take into account a dependence of the LF on 
morphological type and environment. Systematic errors (increasing with 
redshift) could also be introduced as a consequence of errors 
in the sample selection function. Such problems are unavoidable when 
dealing with a flux-limited sample.

Therefore as a complementary approach we have 
reconstructed the density field using a volume-limited subsample
of galaxies. This approach overcomes all the above limitations 
and gives us the possibility to test the robustness of our results 
against different modelling strategies. The price to be paid
is obviously the much smaller number of galaxies; we also neglect
possible effects due to the evolution of the LF, particularly
its faint end: this means that clumps identified in 
the density field of luminous galaxies could have a different density 
of fainter galaxies as a function of redshift.

The advantage is that the two approaches suffer from different
limitations, and obtaining consistent results with both of them
allows us to derive more robust conclusions.

We will now discuss in some detail how we computed the three functions  
$\Phi(m)$, $\zeta(z,m)$ and  $\Psi(\alpha,\delta)$.

\subsection{VMMPS/SPOC sampling rate as a function of apparent magnitude}\label{method1} 
  
In the VVDS, as in most redshift surveys, only a fraction of all
galaxies in the photometric sample satisfying the given flux limit
criteria is targeted: nearly $40$\% for 4 passes area, and a lower
fraction for 3, 2 and 1 pass areas (see Paper I).  Furthermore only a
fraction of the targeted objects yields a reliable, \ie quality
flag $\ge$ 2, redshift: $\sim 80$\% (see Paper I).

Since the VVDS targeting strategy is optimized to maximize the number
of slits on the sky, the selection of faint objects is systematically
favoured \citep{bottini2005, pollo2005, ilbert2005}. As a consequence,
the final spectroscopic sample is slightly biased with respect to the
photometric one, at the bright magnitude end. We thus compute the
correcting function $\Phi(m)$ as the ratio of the distribution of the
magnitudes of targeted objects in all VVDS-02h data to the
distribution of the magnitudes of photometric catalogue.

\subsection{The Redshift Sampling Rate as a function of apparent magnitude and redshift}\label{method2} 

The function $\zeta(z,m)$ provides a correction for two effects: the
progressive degradation toward fainter magnitudes of our ability of
measuring a redshift and the presence of redshift ranges where the
number and strength of identifiable spectral features is scarce.  To
determine this function we followed the approach outlined in
\citet{ilbert2005}, and we refer the reader to that paper for details.

Here is enough to mention that the weights introduced by this function
are determined by comparing the photometric redshift distribution of
all targeted galaxies with the spectroscopic redshift distribution of
all high quality flag galaxies, \ie quality flag$\ge$2. This
comparison is done in the subset area of the VVDS-02h field where J
and K photometry is available \citep{iovino2005}, as the availability
of near-infrared photometry improves the robustness of photometric
redshift estimates.

\subsection{Angular sampling rate}\label{method3} 

The function $\Psi(\alpha,\delta)$ further modulates the sampling rate
defined by the two previous functions. Its purpose is to make
allowance for the number of passes performed by VIMOS in the
$(\alpha,\delta)$ region considered.

This function was calculated in two steps.  In a grid of step size
$s_{step}= 30\arcsec$ in Right Ascension and Declination over the full
VVDS-02h field, we computed in squares of size $s_{box}= 7\arcmin$
(which roughly corresponds to the dimensions of a VIMOS quadrant) the
ratio of the number of objects with a reliable redshift to the number
of potential targets in the same area. This ratio provides for each
grid point the global sampling rate of the VVDS, irrespective of
magnitude and redshift. $\Psi(\alpha,\delta)$ was then obtained by
normalizing to unity the mean value of this ratio over the full 
$\alpha,\delta$ range of the survey.

\begin{figure}
\centering
\includegraphics[width=9cm]{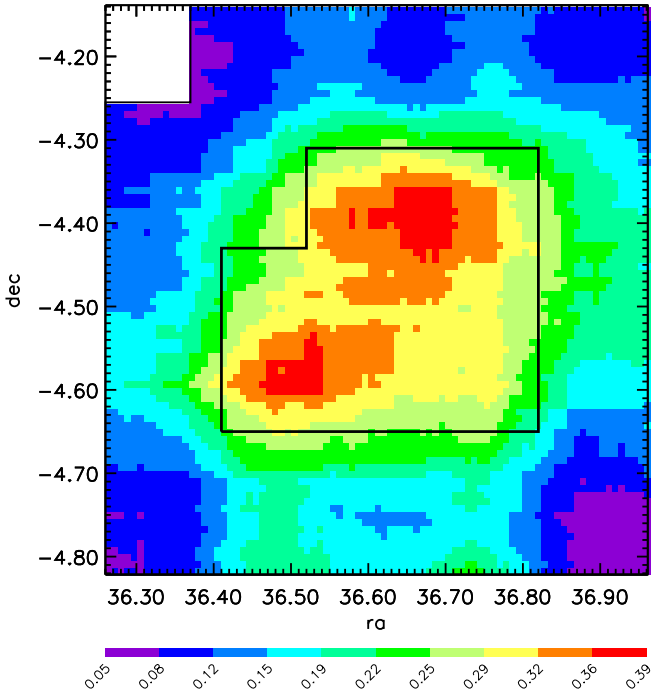}
\caption{The function $\Psi(\alpha,\delta)$ over the full VVDS-02h
field, before normalization. The grid used had steps 
$s_{step}= 30\arcsec$ in Right Ascension and Declination, and for each 
grid position we used squares of size $s_{box}= 7\arcmin$  to estimate 
$\Psi(\alpha,\delta)$.  Both the central area covered by 4 VIMOS passes 
(highlighted by the central continuous line) and the missing N-W corner 
are clearly visible.}
\label{map}
\end{figure}

Figure \ref{map} shows in colour-scale the function
$\Psi(\alpha,\delta)$ before normalization over the full VVDS-02h
field. The central area covered by 4 VIMOS passes is clearly visible,
and corresponds to a sampling rate of 33\%. In other words in the
VVDS-02h-4P area, and down to our selection limit $I \leq 24.0$, on
the mean one galaxy out of three gets a reliable redshift. The
slightly unevenness in the coverage within the VVDS-02h-4P area is due
to some lower quality quadrants. The sampling rate declines going to
region covered by 3, 2 and 1 VIMOS passes. Also the missing N-W corner
is clearly visible.

We explored smaller values for $s_{box}$, down to 1\arcmin, without
any significant improvement of the results obtained (see later
paragraph \ref{simul}). Our final choice for $s_{box}$, being
comparable with the size of a VIMOS quadrant
(7\arcmin$\times$8\arcmin), enables us to take into account the
presence of missing or poorer quality quadrants within a pointing.  In
our computation of $\Psi(\alpha,\delta)$ we suitably adjusted
$s_{box}$ for grid positions near the field boundaries to avoid
introducing spurious border effects.

\subsection{Boundary effects}\label{method4} 

The first-epoch VVDS-02h data sample extends over a square area of
0.7$\times$0.7 $deg^2$ except for a small corner missing, the
North-West one (see Figure \ref{map}).

The effect of the presence of edges is to surreptitiously lower the
measured density. To take into account this problem we scaled the
densities measured around each galaxy dividing by the fraction of the
volume of the filter contained within the survey borders.  Such a
correction can be quite large, especially for galaxies located near to
the survey borders, for example those lying at the corners in Right
Ascension and Declination of our survey layout. To minimize this
correction, and considering that the border regions of the survey are
those with the lowest sampling rate, we introduced a further trimming:
in all our plots we considered only galaxies in positions such that at
least 50\% of the volume of the 3-D gaussian used to define the density
contrast lies inside the volume surveyed. These galaxies, for
R=5 \hpc (R=8 \hpc), amount to 10\% (30\%) of our sample down to
$(M_B-5 \log h) \leq 19.0$.  

Scattered through the entire survey field there are further small
spurious voids due to masking in the photometric catalogue of areas
contaminated by bright stars and diffraction spikes (for a total
masked area of less than 10\% of the total area).  Our simulations
show that neglecting these spurious voids has no significant impact on
our ability to reconstruct the underlying density contrast field.

\section{Estimating Reconstruction Systematics Using Mock catalogues}\label{simul}

We made extensive use of simulations in order to explore the redshift
ranges and smoothing length scales R over which our density
reconstruction scheme (cfr. equation \ref{defrg}) is not affected by
the specific VVDS observational constraints. These include intrinsic
limitations in recovering real space positions of galaxies (peculiar
velocities contaminations, spectroscopic accuracy...), survey
geometrical constraints, sampling and instrumental selections effects.

To this purpose we used mock catalogues extracted from GalICS (Galaxies
in Cosmological Simulations) which is a numerical model of
hierarchical galaxy formation that combines cosmological simulations
of dark matter with semi-analytic prescriptions for galaxy formation
\citep{Hatton2003}.  Thanks to the implementation of the Mock Map
Facility \citep[MoMaF,][]{Blaizot2005} we converted the 3D mock
catalogues into 2D sky images, and handled the 2D projection of the
simulation as a pseudo-real imaging survey. 

Using an approach similar to that described in \citet{pollo2005} and
in \citet{marinoni2005} we constructed VVDS-like mocks which
reproduce the survey angular extension, volume, flux constraints and
spectroscopic resolution. We refer to these samples, which one would
ideally obtain by observing with 100\% sampling rate the VVDS-02h
field, as the {\it parent} catalogues.  These {\it parent} catalogues
are unaffected by the presence of boundaries (they have an angular 
extension much wider than the VVDS-02h field) and they do not contain
masked areas.

To each of these parent catalogues we applied the various instrumental
selection effects and the VVDS observing strategy, including the same
geometrical pattern of excluded regions with which we avoided to
survey sky areas contaminated by the presence of bright stars or
photometric defects, the same target selection procedure and the same
magnitude-distribution of failures in redshift measurements (see paper
I).  In this way we closely matched the actual characteristics of the
VVDS-02h field as observed, and the resulting catalogues are called
{\it observed} catalogues.

Finally, we also constructed catalogues in {\it real space}, that is
catalogues with the same objects as the {\it parent} catalogues, but
without implementing the effects of large scale streaming motions and
measurement errors in the redshift estimate.

The density contrast was therefore reconstructed according to the
prescriptions described in section \ref{method} for parent
($\delp({\bf r},R)$), real space ($\delr({\bf r},R)$) and observed
catalogues ($\delo({\bf r},R)$).

\begin{figure}
\centering
\includegraphics[width=9cm]{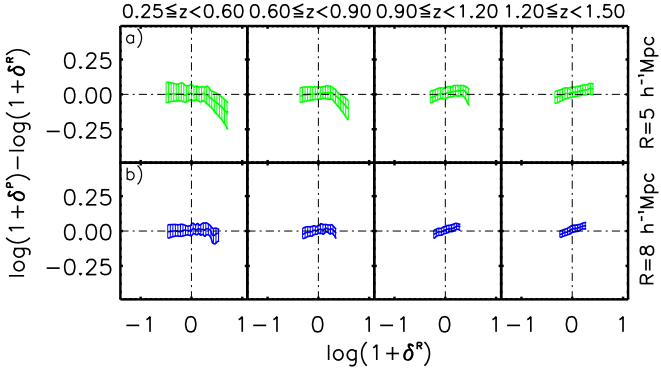}
\caption{The difference between over-densities reconstructed in
redshift space ($\delta^P$) and in real space ($\delta^R$) is plotted
as a function of the real space density contrast.  Results represent 
the average of 50 pseudo independents GalICS mocks.  Columns are for
four redshift bins ($0.25< \z \leq 0.6, 0.6 \leq \z \leq 0.9, 0.9 \leq
\z \leq 1.2$ and $1.2 \leq \z \leq 1.5$) and rows for two different
values of the smoothing scale: R= 5(/8) \hpc from top to bottom.
Central solid lines represent continuous median values computed in 
equipopulated bins, and the boundaries of shaded areas show the 
$25^{th}$ and the $75^{th}$ percentiles.}
\label{simul1}
\end{figure}

\begin{figure}
\centering
\includegraphics[width=9cm]{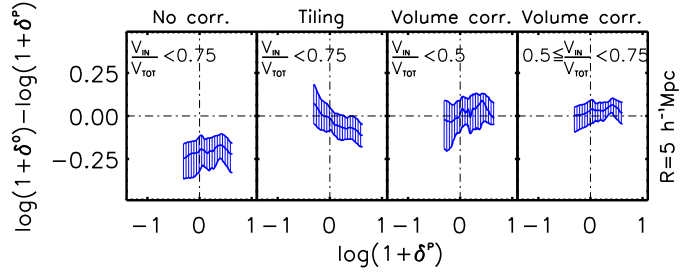}
\caption{The difference between over-densities reconstructed in the
observed ($\delta^O$) and parent ($\delta^P$) catalogue is plotted as
a function of the parent density contrast $\delp({\bf r},R)$.  The
four panels correspond to different strategies of corrections for
border effects, as indicated by the label on the top. $V_{in}/V_{tot}$
in each panel shows the limit on the fraction of the volume of the
gaussian filter F contained within the survey borders for each of the
galaxies plotted. Central solid lines represent continuous median
values computed in equipopulated bins, and the boundaries of shaded
areas show the $25^{th}$ and the $75^{th}$ percentiles. See text for
more details.}
\label{borders}
\end{figure}

\begin{figure}
\centering
\includegraphics[width=9cm]{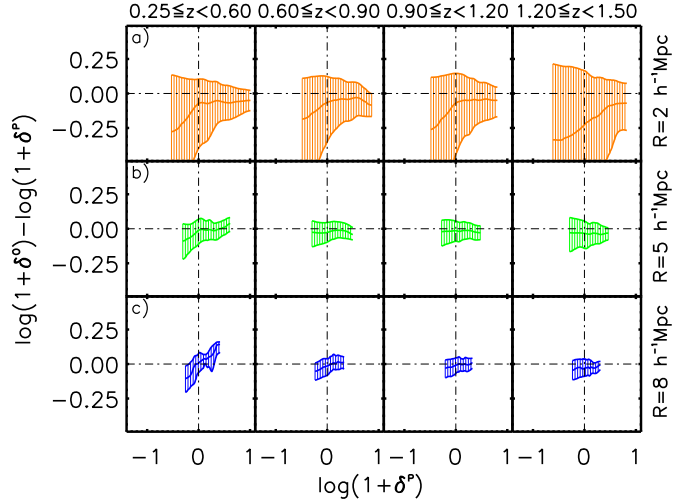}
\caption{The difference between over-densities reconstructed in the
observed and parent catalogue is plotted as a function of the parent
density contrast $\delp({\bf r},R)$.  Columns are for four redshift
bins ($0.25< \z \leq 0.6, 0.6 \leq \z \leq 0.9, 0.9 \leq \z \leq 1.2$
and $1.2 \leq \z \leq 1.5$) and rows for three different values of R:
2, 5, 8 \hpc from top to bottom. Central solid lines represent
continuous median values computed in equipopulated bins, and the
boundaries of shaded areas show the $25^{th}$ and the $75^{th}$
percentiles. Results represent the average of 50 pseudo-independent
GalICS mocks.}
\label{simul2}
\end{figure}

\begin{figure}
\centering
\includegraphics[width=9cm]{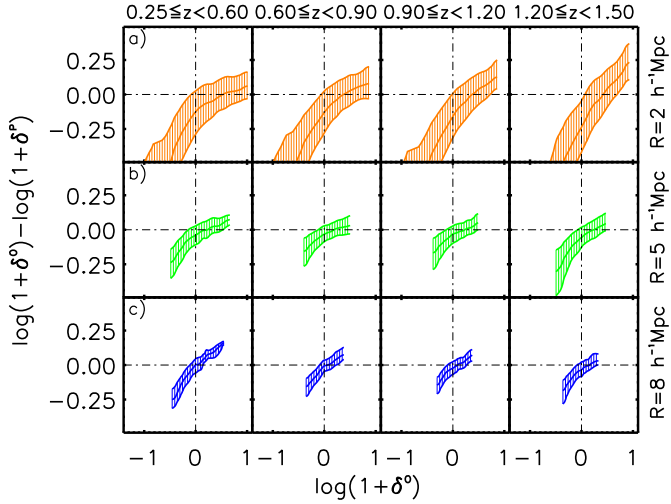}
\caption{As in Figure \ref{simul2}, but here the difference between
over-densities reconstructed in the observed and parent catalogue is
plotted as a function of the observed variable $\delo({\bf
r},R)$.}
\label{simul3}
\end{figure}

Our comparison strategy is twofold. First we check to what extent
non-cosmological {\it kinematical} effects such as galaxy peculiar
velocities and spectroscopic random errors in redshift measurements
smear out the galaxy cosmological redshift and hamper our ability to
recover real space galaxy positions and small scale environmental
densities.  Second we test how {\it geometrical} artifacts introduced
by the specific VVDS target selection strategy and its sparseness
(only 1 galaxy over three at the VVDS magnitude depth has a measured
redshift in the four passes area) degrade the underlying signal.

There are a few obvious guidelines in selecting the ranges of
plausible parameters for reliable density reconstruction.  First of
all, one may expect to exclude smoothing lengths $R << 4$ \hpcv
i.e. much smaller than the mean inter-particle separation of our
sample, which is 4.4 \hpc at the peak of the observed redshift
distribution. We tested, for example, that on these scales a
substantial fraction of galaxies that are classified as isolated (\ie
with no observed neighbour in the reference volume considered) in the
observed catalogue is not constituted by truly isolated galaxies in
the parent catalogue, but by galaxies populating over-dense regions
($\delta\geq0$) for which the density field reconstruction fails.  We
also excluded from the analysis regions with redshift greater than $z
= 1.5$, where the sample becomes too sparse and the mean inter-galaxy
separation too large. Similarly, the lower redshift limit \z$_t$ is
set by imposing that the transversal dimension L of the field 
be L$(z_t)>$R.  As an example, for a Gaussian
window of size R=8 \hpcv we have \z$_t\sim$ 0.25.

Figure \ref{simul1} shows the difference between densities
reconstructed in the parent catalogue ($\delp$) and in real space
($\delr$) versus the real space density contrast. We see that on
scales $R=5$ \hpcv the parent catalogue (that is redshift-space)
densities are systematically underestimated in over-dense
regions. This is because, on small scales, non-linear structures
observed in redshift space are smeared out along the line of sight
(the so-called {\it Finger of God} effect).  On larger scales we
expect the opposite (Kaiser effect), i.e. that the infalling pattern
towards over-dense regions spuriously enhances the density contrast
recovered in redshift space. Effectively, the transition between these
two different regimes becomes appreciable on scales R=8\hpcp

For our study as a function of environment, it is of fundamental
importance to differentiate in a robust way between over- and
under-dense regions, and Figure \ref{simul1} shows that under-dense
regions are safely recovered in redshift space. Moreover, the small
amount of the underestimation in over-dense regions (for $R\geq5$
\hpc) guarantees that there is no fictitious percolation of high
density into low density environments and vice-versa.
 
A possible concern in our density reconstruction scheme could be
on systematics introduced by our chosen strategy to correct for survey
borders. Our survey is not a pencil beam survey in the original sense,
but still its transverse dimensions are much smaller than its
dimension along the redshift axis. Figure \ref{borders} shows the
difference between observed and parent density contrast $\delta$ as a
function of the parent $\delp$ for three possible strategies of border
effects corrections in an observed catalogue that is like the parent
except for the presence of VVDS-like borders. 
In this plot we show only galaxies such that the fraction $V_{IN}/V_{TOT}$
of the volume of the gaussian filter F contained within the survey borders
is less than a fixed value, as indicated in each panel.

In the first panel no correction for borders is implemented and, as
expected, a systematic underestimate of the parent density is clearly
visible for border galaxies. In the second panel we show the results
obtained when applying a tiling correction. 
We tiled around the
observed VVDS-02h field 8 replicas of the VVDS-02h field itself (after
adding a smaller tile to cover the missing N-W corner, see Figure \ref{map}). 
We tested the robustness of the tiling correction as follows:
we tiled the 
mocks simulating the VVDS-02h field and compared the recovered densities 
with the underling ``true'' densities of the parent simulations 
which has a larger extension than the VVDS field.
The tiling strategy allows a more reliable reconstruction 
of the environment, at least in a statistical way, for 
galaxies near the edges of the survey. Anyway, still a non negligible
underestimate of overdense regions and overestimate of underdense
regions is visible. This is due to the fact that the tiling smeares
the original densitities by adding for the border objects a random
density to the actual density in the parent catalogue. Finally the
last two panels show the results of a volume correction scheme. We
correct the density contrast measured around each observed galaxy by a
factor that corresponds to the inverse of the fraction of the volume
of the gaussian filter F, centered on the galaxy in question, inside
the VVDS-like mock survey. By comparing the reconstructed density with the 
``real'' density field of the parent simulation we conclude that 
there is no large, systematic shift between
observed and parent $\delta$, and especially so when excluding all
galaxies such that this correction is higher than 2 as in the last
panel. This is the solution we adopted. 

Finally we turn to the question of assessing how well the observed
galactic environment, reconstructed after applying the whole VVDS
pipeline to simulations, traces the underlying parent over-density
field.  In Figure \ref{simul2} we plot the difference between the
logarithm of the {\it observed} and of {\it parent} over-densities
($\Delta=log(1+\delo({\bf r},R))$-$log(1+\delp({\bf r},R))$) with
respect to the logarithm of the density contrast $log(1+\delp({\bf
r},R))$ as a function of redshift. Rows refer to three different
values of R: 2, 5, 8 \hpc from top to bottom. In this plot and in
all subsequent plots we followed the recipe discussed in the
subsection \ref{method4} for border corrections and used only galaxies
located in a position with respect to survey borders such that less
than 50\% of the volume of the 3-D gaussian centered on them and used
to define the density contrast lies outside the volume surveyed.

In this way we can check for possible systematics affecting the
observed over-density: \eg a systematic over/underestimate of density
contrast introduced by the wrong functions in equation \ref{defrg}, or
by non-trivial border effects introduced by masking of some areas of
the observed catalogue. It is evident how badly we reconstruct the
density field on scales as small as 2 \hpc, as expected from the
previous discussion.  For $R \geq 5$\hpc, there is no evidence of
systematic biases in the reconstruction of the over-density
distribution. This conclusion holds irrespective of the particular
redshift range investigated in the interval $0.25<z<1.5$. 

However, using real data, we have only access to the information
contained in the observed over-density. The conditional distribution
of $\delta$ given the observed over-density $\delo$
($P(\Delta|\delo))$ is expected to be different from
$P(\Delta|\delp)$.  Therefore, we analyzed the difference between
observed and parent over-densities as a function of $\delo({\bf r},R)$
in Figure \ref{simul3}.  In this way we can directly assess, in terms
of observed quantities, in which over-density ranges the VVDS
environment catalogue is reliable.  We see that, for $R < 5$\hpc, the
observed over-density field underestimates the underlying parent
density contrast in under-dense regions and overestimates it in
over-dense ones. Only for $R \geq 5$\hpc the observed over-densities
trace the underlying true distribution in a fair way.

In short, the results of simulated VVDS observations presented in this
section show that, on scales R $\ge$ 5 \hpcv 3D over-densities are
essentially free from selection systematics at least for what concerns
the unbiased identification of galaxy environments in both low and
high density regions. Obviously, the representativeness of the
measured environments with respect to the ``universal'' one is a
different question. Since the volume probed is still restricted to one
survey field, the dynamical range of the recovered over-densities may
be affected by cosmic variance \citep[\eg][]{ilbert2006zphot}. 
From now on we will use 3D over-densities as measured
with a filter of size R = 5 \hpc.

\section{Results}\label{results}

In this section we present our results on the dependence of galaxy
colours from local density, luminosity and redshift.  As
\citet{blanton2005} among other authors have shown, colour is the
property that best correlates, together with luminosity, with local
environment. For this analysis we use rest-frame ($u^{*}-g'$) colours,
uncorrected for dust absorption, derived from rest-frame AB absolute
magnitudes as computed in the $u^{*}$ and $g'$ CFHTLS-MEGACAM photometric
system. The strategy adopted to derive rest-frame absolute magnitudes
is described in detail in \citet{ilbert2005}, and we refer the reader
to that paper for details. The CFHTLS-MEGACAM photometric system
has been designed to match the SDSS filters as closely as possible,
with the only exception of the $u^{*}$ filter, that is slightly wider
than the SDSS $u'$ filter (see \citealp{ilbert2006zphot}).

\begin{figure*}
\centering \includegraphics[width=15cm]{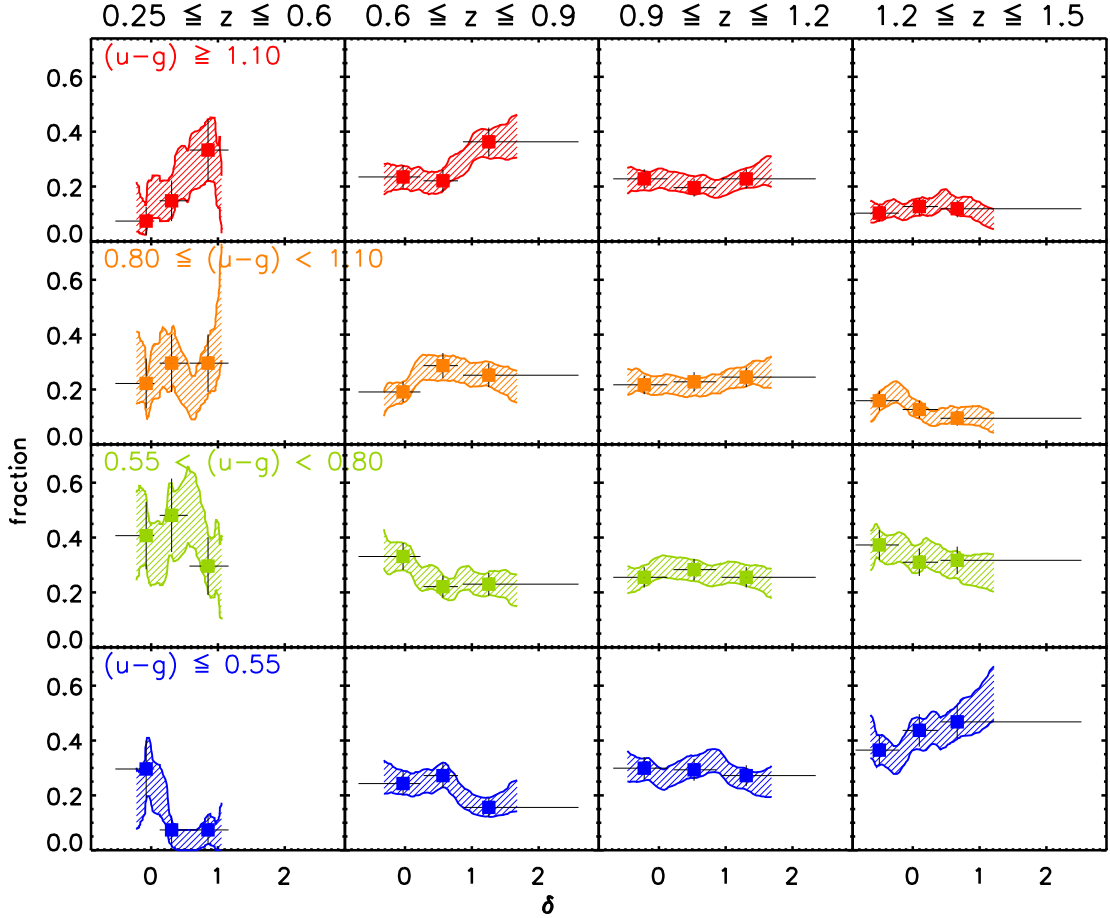}
\caption{ The fraction of galaxies with luminosities greater than
$(M_{B}-5\log h) < -20$ and with rest-frame colours $(u^{*}-g') \geq 1.1$,
$0.80 \leq (u^{*}-g') < 1.1$, $0.55 < (u^{*}-g') <0.8$ and 
$(u^{*}-g') \leq 0.55$ (from top to bottom) is plotted as a function of 
environment ($\delta$ on R = 5 \hpc) in four different redshift intervals 
(from left to right, as indicated on the top).  The horizontal bars 
indicate the amplitude of the bins in $\delta$ (i.e. the range spanned by 
the lower 5\% and upper 95\% percentile of the objects contained in each 
bin) while vertical error-bars represent a Poissonian $1 \sigma$
uncertainty. The shaded areas are obtained by smoothing the red (blue)
fraction with an adaptive sliding box containing the same number of
objects in each bin as the points marked explicitly.}
\label{figure1}
\end{figure*}

We choose the ($u^{*}-g'$) colour because it brackets the Balmer break
and is therefore particularly sensitive to galaxy properties (\eg age,
recent star formation and metallicity variations of the stellar
population).  Another advantage of this colour choice is that, given
the range of filters available to our survey, rest-frame absolute
magnitudes are better reconstructed (out to $z \sim 1.5$) in the $u^{*}$
and $g'$ bands than in other redder filters \citep{ilbert2005}.

\subsection{The colour-density relation: redshift and luminosity dependence}\label{results1} 

We empirically defined 4 different galaxy ``colour-types'' on the
basis of the following rest-frame colour criteria: $(u^{*}-g') \geq 1.1$,
$0.80 \leq (u^{*}-g') < 1.1$, $0.55 < (u^{*}-g') <0.8$ and 
$(u^{*}-g') \leq 0.55$. Note that our reddest and bluest colour classes 
roughly correspond to the two colour peaks visible in the bimodal colour 
distribution shown in Figure \ref{figure4}.

In Figure \ref{figure1} we plot the fraction of galaxies of each
colour-type and with luminosities greater than $(M_{B}-5\log h) < -20$
as a function of the environment (\ie the density contrast
$\delta$). The colour-density relation is portrayed at four different
cosmic epochs. This figure shows one of the key observational results
of our investigation: the significant redshift dependence of the
colour-density relationship. In the lowest redshift bin ($0.25 \leq z
\leq 0.60$), the bluest (reddest) galaxies are preferentially located
in low (high) density regions, the trend changing smoothly through
intermediate colours. This trend, which is reminiscent of the
well-known local morphology-density relationship, progressively
disappears and possibly reverses in the highest redshift bin ($1.20
\leq z \leq 1.50$).

On top of this trend, another important key feature is evident: the
strong evolution in the mean fraction of the bluest galaxies as a
function of cosmic time.  The relative abundance of these objects
increases with increasing redshift, in agreement with what found by
other redshift surveys of the deep universe
\citep[e.g.,][]{lilly95,lin99}.  However Figure \ref{figure1} adds an
important element to the picture: the fraction of blue galaxies
increases with increasing redshift not only in rich environments but
also in under-dense regions (see last row of Figure
\ref{figure1}).

\begin{figure*}
\centering
\includegraphics[width=15cm]{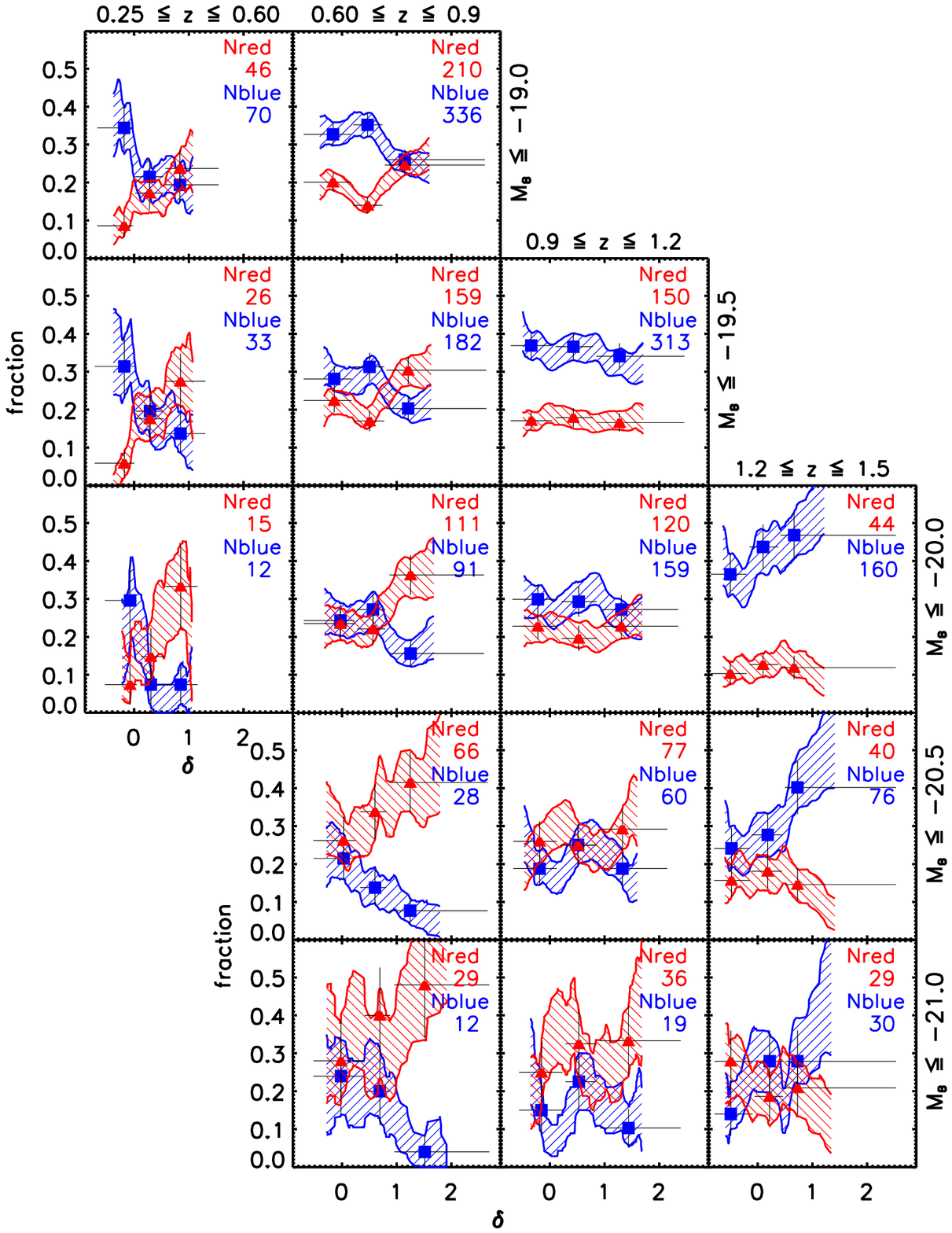}
\caption{ The fraction of the reddest ($(u^{*}-g') \geq 1.10$, triangles)
 and bluest ($(u^{*}-g') \leq 0.55$, squares) galaxies is plotted as a
 function of the density contrast $\delta$ in different redshift
 intervals (columns, as indicated on top) and for different absolute
 luminosity thresholds (rows, as indicated on the right).  The
 horizontal bars indicate the amplitude of the bins in
 $\delta$ (i.e. the range spanned by the lower 5\% and upper 95\%
 percentile of the objects contained in each bin, the marked points
 being located at the median value) while vertical error-bars
 represent a Poissonian $1 \sigma$ uncertainty. The shaded areas are
 obtained by smoothing the reddest(bluest) fraction with an adaptive
 sliding box containing the same number of objects in each bin as the
 points marked explicitly. The number of red and blue galaxies in 
 each redshift and luminosity bin is explicitly indicated in the 
 corresponding panel.}
\label{figure2}
\end{figure*}

\begin{figure*}
\centering
\includegraphics[width=15cm]{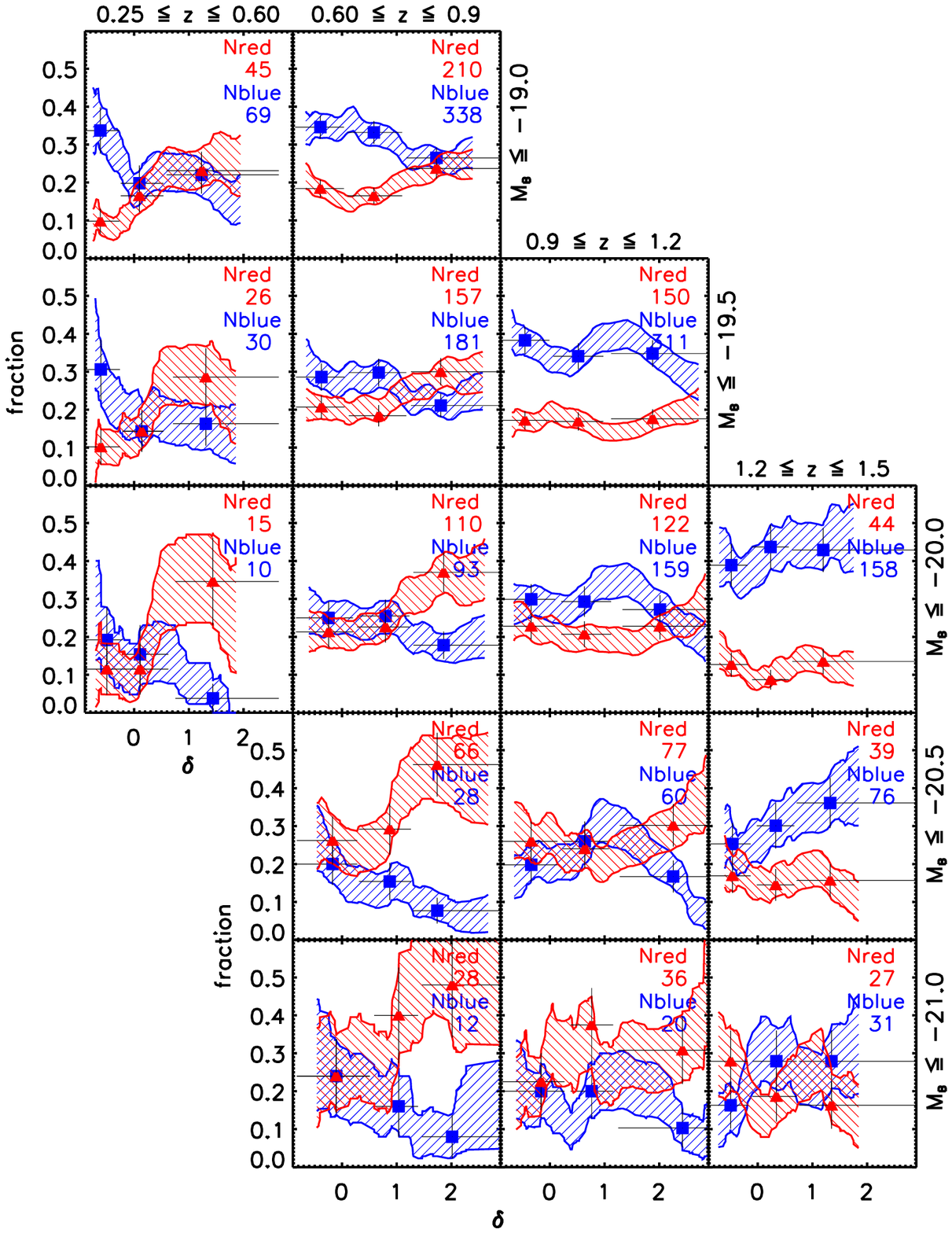}
\caption{ As in Figure \ref{figure2}, but computing local densities
using only galaxies brighter than $(M_B-5 \log h) = -20.0$. See text
for details.}
\label{figure2VL}
\end{figure*}

We also explored the combined dependence of the colour-density
relation on redshift and luminosity. To this purpose, we selected
different samples of galaxies, using as luminosity thresholds the
values $(M_B-5 \log h) \leq -19.0, -19.5, -20.0, -20.5, -21.0$
respectively. For each of these samples the fractions of the reddest
($(u^{*}-g') \geq 1.10$) and bluest ($(u^{*}-g') \leq 0.55$) galaxies 
are shown in Figure \ref{figure2} as a function of $\delta$ in four 
different redshift bins. 

\begin{figure*}
\centering \includegraphics[width=15cm]{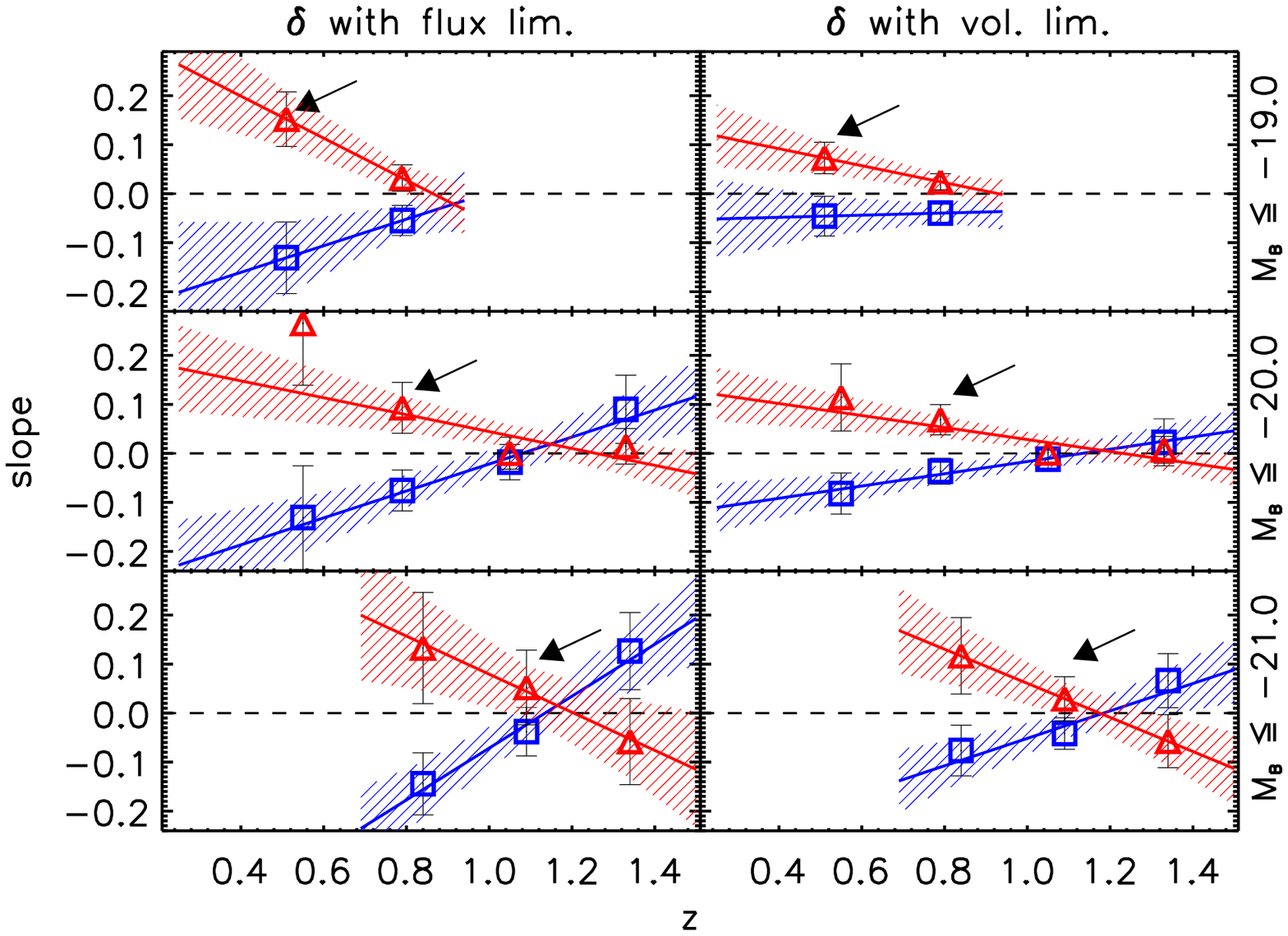} 
\caption{ Best fit slopes (and their associated 1 $\sigma$ error bars) 
of the fraction of reddest ($(u^{*}-g') \geq 1.10$, triangles) and 
bluest ($(u^{*}-g') \leq 0.55$, squares) galaxies as a function of 
$\delta$ for the four different redshift bins of Figure \ref{figure2} 
(left column) and of  Figure \ref{figure2VL} (right column). From top 
to bottom different limits in absolute magnitude are considered, as 
indicated on the vertical label on the right. In each panel the straight 
lines (and the shaded error bar area associated) are the result of the 
linear fit to the points shown.  The black arrows indicate the redshift 
bin where the colour--density relationship, as we know it in the local 
universe, appears for the first time. }
\label{figure2bis}  
\end{figure*}

Note that a) the survey flux limit ($I=24$) forces us to restrict the
analysis of the fainter samples only to the lower redshift bins, b)
the limited area of the survey prevents us to collect adequate signal
from bright objects in the lowest redshift bin, because the volume
explored at low redshift is not large enough to sample the
exponentially decreasing bright end of the luminosity function.

While all the samples used in the first three redshift bins of Figure
\ref{figure2} are indeed samples purely limited in absolute magnitude,
the samples in the last redshift bin could be partially affected by
colour incompleteness. Our survey is flux-limited at $I\leq 24$ and
the $I$-band corresponds to $B$-band rest frame wavelength at redshift
$z\sim0.8$. For redshifts greater than this value the absolute
magnitude cut-off that selects our samples will be a function of the
colour of the galaxy population considered - see \eg Figure 2 in
\citet{ilbert2004}. In particular while at $ z\leq1.2$ we can safely
assume to be complete down to $(M_B-5 \log h)\sim -20.0$ irrespective
of galaxy colours, when moving to the last redshift bin $1.2\leq z\leq
1.5$ only the sample limited at $(M_B-5 \log h)\sim -21.0$ is complete
for galaxies whose colour is as red as that of evolved ellipticals
observed in the local universe.

Therefore, the overall normalization of the $(M_B-5 \log h) \leq
-20.0$ and $\leq -20.5$ reddest samples in the last redshift bin
should be considered as a lower limit.

As an additional test on the density estimates obtained with equation
\ref{defrg}, we computed local densities also using a volume limited
sample ($(M_B-5 \log h) \leq -20.0$).  

By using as the population to define the density contrast $\delta$
only galaxies brighter than a fixed $B$-band absolute magnitude, we
can drop the redshift dependent selection function $S(r,
\mathcal{M}^c)$ in equation \ref{defrg}.  This way the noise of
the density estimate does not depend on redshift nor on the luminosity
function, whose estrapolation at high redshifts is plagued by
uncertainties. Futhermore we avoid introducing, especially at high
redshifts, an artificial homogeneity in the galaxy distribution, as
progressively brighter galaxies are used to trace the space
distribution of the full galaxy population.  The results obtained with
this recipe are displayed in Figure \ref{figure2VL}, that shows the same
trends as in Figure \ref{figure2}, although a bit noisier,
especially in the first redshift bin where the density reconstruction
is based on a much smaller number of galaxies than in Figure \ref{figure2}.

Both Figure \ref{figure2} and \ref{figure2VL} show that not only the
colour segregation weakens as a function of redshift for galaxies of
similar luminosity, but, at a fixed redshift, it strongly depends on
luminosity: for progressively brighter galaxies the colour--density
relationship, as we know it in the local universe, appears at earlier
cosmic times.

To quantify the statistical significance of our findings, we fitted
the points plotted in these Figures with a linear relation ($f = a +
b\delta$, where $f$ is the fraction of red or blue galaxies).  The 1
$\sigma$ error bars obtained by fitting our data well agree with those
obtained with randomization techniques. In this case we randomized
1000 times, for each redshift and luminosity bin, the distribution of
$\delta$ among our galaxies, and for the randomized sample we plotted
the quantities shown in Figure \ref{figure2} and \ref{figure2VL},
testing how often the linear fit to the randomized data would provide
a slope as steep or steeper than the one measured from our data.

Figure \ref{figure2bis} shows the slopes $b$ and the associated 1
$\sigma$ error bars as a function of redshift, for red ($(u^{*}-g') \geq
1.1$, triangles) and blue ($(u^{*}-g') \leq 0.55$, squares) galaxies, for
the three subsamples limited at $(M_B-5 \log h) = -19.0, -20.0, -21.0$
going from top to bottom. Left panel refers to Figure \ref{figure2},
\ie when density contrast is estimated using the full flux limited sample,
while right panel refers to Figure \ref{figure2VL}. The black arrows
indicate the redshift bin where the colour--density relationship, as
we know it in the local universe, appears for the first time.

Considering the results obtained using a flux limited sample to
estimate densities, at low redshift ($0.25 \leq z \leq 0.60$) we find
that for all the subsamples considered red galaxies are preferentially
located in high density regions (positive slope), while blue galaxies
are preferentially located in low density regions (negative
slope). These slopes are different from zero at more than 2 $\sigma$
level for red galaxies and nearly 2 $\sigma$ level for blue
ones, and their relative difference for each panel in the first
column of Figure \ref{figure2}, is significant at the 2.5 - 3
$\sigma$ level. In the next redshift bin ($0.60 \leq z \leq 0.90$),
significant differences (at 2.5 $\sigma$ level) between the
slopes of the fraction of red and blue galaxies as a function of
$\delta$ are present only for galaxies brighter than $(M_B-5 \log h) =
-20.0$. Viceversa for galaxies fainter than $(M_B-5 \log h) = -20.0$
no significant trend with density is seen for the fractions of both
red and blue galaxies. At even higher redshift ($0.90 \leq z \leq
1.20$), some difference between the slopes of the fraction of red and
blue galaxies as a function of $\delta$ is visible only for the very
brightest galaxies ($(M_B-5 \log h) \leq -21.0$), although at a low
level of significance (1 $\sigma$), also because of the small
statistics. Finally in the highest redshift bin explored ($1.20 \leq z
\leq 1.50$), for the very brightest galaxies ($(M_B-5 \log h) \leq
-21.0$), the slope of the fraction of red(blue) galaxies as a function of
$\delta$ is even negative(positive), though at 1(1.6) $\sigma$ level.

When the density contrast is estimated using the $(M_B-5 \log h) \leq
-20.0$ sample the significance of our findings is slightly lower.
This is partially due to the noisier estimates of densities, at least
in the first redshift bins, but the substance of our results does not
change, as one can estimate from Figures \ref{figure2VL} and
\ref{figure2bis}.

\subsection{The colour-magnitude diagram: redshift and density dependence}\label{bimo}

In this section we explore the evolution of the distribution of
galaxies in the colour-magnitude plane ($u^{*}-g'$) vs. $(M_B-5 \log
h)$ as a function of both redshift and environment, further expanding
the correlations discussed in the previous section.

\begin{figure}
\centering \includegraphics[width=9cm]{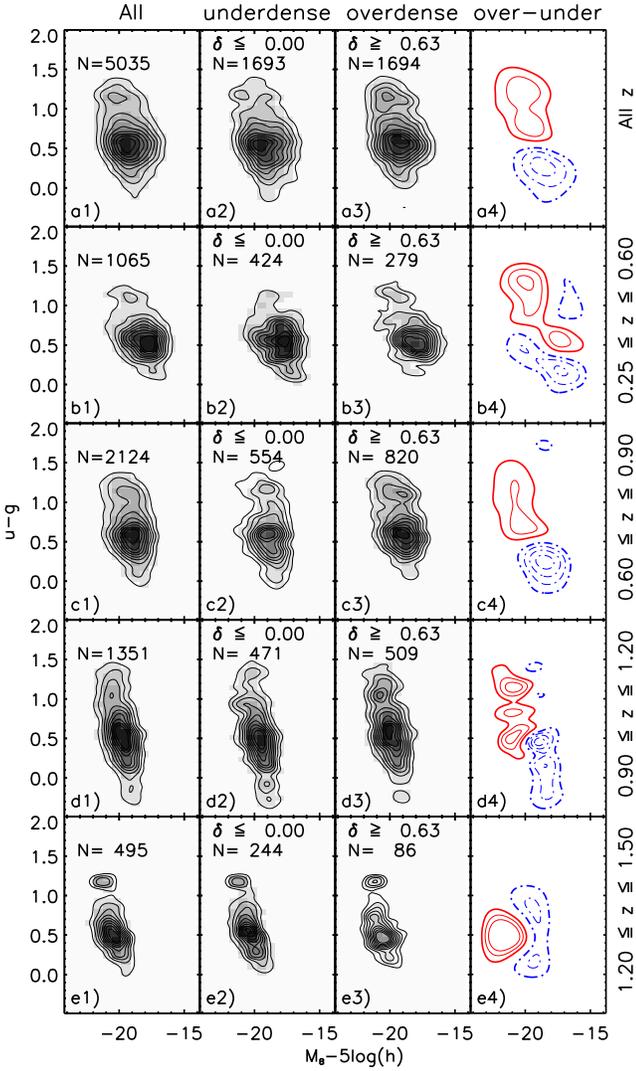}
\caption{The first 3 columns show the isodensity contours of the
distribution of galaxies in the ($u^{*}-g'$) vs. $(M_B-5\log h)$ plane
for different redshift ranges (from top to bottom as indicated on the
right) and for different environments (from left to right as indicated
on top).  The grey scale is normalized, for each panel, to the total
number of objects contained in that panel.  The difference between the
over-dense and under-dense colour-magnitude distributions is shown in
the 4th column.  $1-2-3-\sigma$ levels of significance in the
difference are shown using red continuous lines (positive
differences), and blue dotted lines (negative differences). The
thicker lines correspond to $1\sigma$ level.}
\label{figure4}  
\end{figure}

In Figure \ref{figure4} the first 3 columns show the isodensity
contours of the distribution of galaxies in different redshift ranges
(from top to bottom as indicated on the right) and for different
environments (from left to right as indicated on top).  The difference
between the over-dense and under-dense colour-magnitude distributions
is shown in the fourth column.

The first column shows that the bimodal distribution of galaxies in
colour space, well established in the local universe \citep[e.g.,
][]{Strateva2001}, persists out to the highest redshift investigated
($z\sim 1.5$).  This analysis confirms and extends at higher redshifts
previous results obtained with photometric redshifts out to $z=1$
\citep[e.g., ][]{Bell2004,nuijten2005}.  A more detailed analysis of
the bimodality of our sample is presented in \cite{franzetti2006}.

The second and third columns of Figure \ref{figure4} show that
bimodality survives irrespective of environment out to $z\sim 1.5$.
Beyond the bimodality, it is important to notice that the location of
the colour `gap' between the red and blue peaks appears to be roughly
constant and insensitive to environment at all redshifts. This result
justifies {\it a-posteriori} our choice of a fixed colour threshold in
splitting the total sample into a blue and a red subset.

We can discriminate finer environmental dependencies imprinted in the
bimodal colour distribution by plotting the difference between the
over- and under-dense colour-magnitude distributions.  The fourth
column of Figure \ref{figure4} shows that {\it the colour-magnitude
distribution is not universal but strongly depends upon environment.} At
low redshift, and for any luminosity, there is a prominent excess of
red objects in over-dense regions, while under-dense regions are
mostly populated by blue galaxies.  On the other hand, and most
interestingly, moving towards higher redshifts the relative ratio of
the two peaks of the bimodal distribution becomes mostly insensitive
to environment (at $0.9<z<1.2$) with the hint of the development of a
more pronounced peak of blue galaxies in high density regions in the
last redshift bin ($1.2<z<1.5$). Stated differently, the bulk of the
red population in the interval $0.9<z<1.2$ is found to be equally
distributed in different environments, with the brightest red sample,
however, still biased toward high density regions. At higher
redshifts, $1.2<z<1.5$, even the brightest red galaxies are not
preferentially found in over-dense regions, which, instead, become
mostly populated by bright blue objects.

\section{Discussion}\label{discussion}

The most striking result of this study is displayed in Figure
\ref{figure2}: the colour-density relation evolves dramatically as a
function of cosmic time.  While at the remotest epochs explored ($z
\sim 1.5$) even the most luminous red galaxies do not reside
preferentially in high density environments, as cosmic time goes by,
the environmental dependence of galaxy colours progressively builds
up, earlier for brighter galaxies and later for fainter galaxies.  For
example, the faintest red galaxies ($M_B<-19.0 +5 \log h$) are
preferentially located in high density environments only in the
nearest redshift bin investigated, confirming earlier hints in this
direction obtained with photometric redshifts \citep[\eg
][]{kodama,yee}.  Vice versa in the redshift bin $0.9 \leq z \leq 1.2$
all but the most luminous red galaxies, $(M_B- 5 \log h) < -20.5$,
show a flat colour-density relation.  In the highest redshift bin even
the brightest red objects are not preferentially found in rich
environments as suggested by the fact that the slope of the
colour-density relation for red objects turns negative ($1\sigma$
effect).  In this redshift bin there is a suggestion that the
percentage of bluest galaxies increases in the highest density
contrast regions, hinting that in remote look-back times the star
formation activity was higher in high density peaks than in low
density regions, a property reminiscent of a similar characteristic of
Ly-break galaxies \citep{foucaud2003}.  We conclude that at $z\sim 1$
there is evidence of absence of the colour-density relation for medium
luminosity galaxies. Moreover, there are hints that the well
established local trend, which progressively disappears even for the
brightest galaxies in our sample after $z \sim 1.2$, eventually
reverses in the highest redshift bins investigated ($\sim 1\sigma$
effect).

Not only the slope of the colour-density has changed, but also the
overall normalization.  The decrease of the relative fraction of the
bluest galaxies from high redshift to the present day is well
established in literature and can be traced back to the observations
of the increase in the abundance of star-forming galaxies in high
redshift clusters \citep[see][]{beo}.  In our study, we find that this
trend holds true {\it also in low density environments}.

Figures \ref{figure2} and \ref{figure2VL} show that the fraction of
the bluest galaxies brighter than $(M_B-5\log h)<-20$ which inhabit
low density regions at $1.2 \leq z \leq 1.5$ decreases on the mean as
cosmic time goes by.  On the other hand, over the same redshift range,
the relative abundance of blue objects has changed by nearly one order
of magnitude in over-dense environments.  Similarly the reddest
galaxies experience a faster increase with cosmic-time in over-dense
regions than in under-dense regions. This result indicates that the
mechanisms governing galaxy formation and evolution operate with
different timescales in different environments.

We can interpret these findings by making some simplifying
hypothesis. Let's assume, to first order, that the adopted colour
classes are a proxy for different star formation histories, bluer
galaxies having experienced relatively recent star formation. In this
case, the observed strong time dependence of the colour-density
relation implies that star formation is differentially suppressed in
high and low density regions. For galaxies of similar luminosity the
drop in star formation rate occurred earlier in higher density
environments, resulting in the red excess observed at present epoch,
and progressively later in lower density environments, \ie in the
field, where a larger blue component is still observed. This result
suggests that some environment driven mechanism may be at work. The
drop in star formation is also a function of luminosity (and therefore
probably mass): truncation mechanisms are more efficient in brighter
systems than in fainter ones.

By further assuming that it is empirically possible to define early
and late type galaxies in an unbiased, model-independent way by
exploiting the bimodality of the galaxy colour distribution out to the
highest redshift investigated \citep[e.g., ][]{Bell2004} our findings
would imply a change in the morphology-density relation
\citep{dressler1980} for galaxies brighter than $M^{*}_B(z=0)$
\citep[$\sim -20+5\log h$, see ][]{ilbert2005} at $z\sim 1$ and for
brighter ones at $z\sim 1.5$.

Even if there are evidences that most colour selected red galaxies are
dominated by an old stellar population from $z=0$ \citep[e.g.,
][]{Strateva2001} up to $z\sim 1$ \citep{Bell2004b}, the situation at
larger redshift is unclear and a substantial fraction of red galaxies
could be dusty starbursts \citep[e.g., ][]{cimatti03}.  Therefore, we
caution that our results should be interpreted as an upper limit on
the distribution of red passive objects. If red dusty starburst
galaxies inhabit high density regions, then the deficit of red old
objects in high density environments at high redshift should be even
stronger than that estimated in our analysis.

From an observational side our analysis well agrees with the so called
{\it downsizing} scenario, first suggested by \citet{gavazzi96} and
\citet{cowie96}, but modified to take into account the observed
environmental dependence. According to our observations, star
formation activity is not only progressively shifted to smaller
systems, but also from higher to lower density environments.

This result agrees remarkably well with our findings (obtained with
the same sample) about the significant evolution of galaxy biasing out
to $z\sim 1.5$ \citep{marinoni2005}.  In that study we showed that we
live in a special epoch in which the distribution of galaxies with
$M_B-5 \log h\sim-20$ traces the underlying mass distribution on
scales $R \geq 5$\hpc while, in the past, the two fields were
progressively dissimilar and the relative biasing higher. In other
words, while at high redshift bright galaxies formed preferentially in
the high matter-density peaks, as the Universe ages, galaxy formation
begins to take place also in lower density environments.  This result
on biasing evolution provides a simple and intuitive way to introduce
environment in the original downsizing picture: brighter galaxies
start forming stars earlier {\it and} preferentially in higher density
environments.

From a theoretical perspective, in models of hierarchical galaxy
formation \citep{kauffmann93, somerville99, cole00}, it is assumed
that massive galaxies, which accreted earlier and in a biased way with
respect to the underlying matter density field, have their hot gas
reservoir depleted, which results in a premature truncation of the
star formation activity relative to field galaxies.  Our findings
about galaxy biased formation coupled with simple assumptions for the
faster gradual decline of the star formation activity of galaxies in
dense environments are able to explain, in a qualitative way, the
observed evolution of the colour-density relation, i.e. the faster
progressive building up of bright red galaxies in high density
environments and the slower evolution for the fainter galaxy
population.

Several physical processes have been proposed that may account for a
consumption/expulsion/evaporation of gas in high density environments:
ram pressure stripping \citep{gunn_gott1972}, galaxy-galaxy merging
\citep{toomre1972}, strangulation \citep{larson1980} and harassment
\citep{moore1996}.  However the scale over which the over-density
field of the deep universe is reconstructed in this paper, prevents us
from concluding on the possible local causes of the observed evolution
in the colour-density relation. In other words, we cannot discriminate
if the mechanism responsible for the differential evolution is acting
also at large distance from the high density cluster core regions
\citep[e.g., ][]{balogh97}.

The strengthening of the colour-density relation as a function of
cosmic time implies that the colour distribution has been tightly
coupled to the underlying density field at least over the past 9
Gyr. The effects of this coupling are evident in Figure \ref{figure4}
where we show the different evolution of the colour-magnitude
distribution in high and low density environments.  The early epoch
flatness of the colour-density relation causes the bimodal colour
distribution in high density regions to mirror the one in poor
environments. However, as time goes by, the colour-density relation
strengthens and the bimodal distribution gradually develops the
present-day asymmetry between a red peak more prominent in high
density environments and a blue one mostly contributed by field
galaxies.

Besides evolution in the gradient and amplitude of the colour-magnitude
distribution, one may however notice a few interesting features which
are stable across different cosmic epochs.  The peak position of the
red population remains nearly unchanged in both over- and under-dense
environments out to $z \sim 1.5$. This implies that {\it a population
of red objects of bright luminosities and in different environments is
already well evolved by redshift 1.5} (see also \citealp{lefevre_lam2006}, 
in preparation). This finding may be easily,
and perhaps most naturally, interpreted as supporting evidence for a
scenario in which old, massive, quiescent objects were already in
place at redshift 1.5 and have undergone very little evolution since
then.  Figure \ref{figure4} shows another interesting similarity
between the low and high redshift universe: brighter galaxies are
redder both in low and high density environments and this holds true
at all redshifts investigated.  We thus find that the colour-magnitude
relation, which has been well investigated in clusters up to $z \sim
1$ \citep[e.g., ][]{visvanathan1977,bower,holden2004,tanaka}
also applies, irrespective of cosmic epochs, to galaxies
populating very under-dense environments.  

\section{Conclusions}\label{conclusion}  

The size (6582 galaxies with good quality redshifts), depth
($I_{AB}\leq 24$) and redshift sampling rate (20\% on the mean) of
the VVDS-02h deep survey allowed us to reconstruct the 3D galaxy
environment on relatively local scales ($R=5$\hpc) up to redshift
$\z=1.5$ and to study the colour distribution as a function of density,
luminosity and look-back time. 

Environmental studies at high redshift have traditionally focused only
on high density regions (galaxy clusters), and/or have been based on
galaxy position inferred using photometric redshifts.  Our study
represents the first attempt to use a purely flux-limited redshift
survey to explore the primordial appearance of the colour-density and
colour-magnitude diagrams from the densest peaks of the galaxy distribution 
down to very poor environments and faint magnitudes.

We have paid particular attention to calibrate our density
reconstruction scheme, and the extensive simulations presented in this
paper enable us to determine the redshift ranges and smoothing length
scales R over which our environmental estimator is not affected by the
specific VVDS observational constraints. These include intrinsic
limitations in recovering real space positions of galaxies (peculiar
velocities contaminations, spectroscopic accuracy...), survey
geometrical constraints, sampling and instrumental selections effects.
We conclude that we reliably reproduced the underlying {\it real}
galaxy environment on scales $R\ge 5$\hpc out to z=1.5.

Our findings  can be summarised as follows:

a) {\it The colour-density relation shows a dramatic change as a
function of cosmic time}.  While at the lowest redshifts explored we
confirm the existence of a strong colour-density relation, with the
fraction of the red(/blue) galaxies increasing(/decreasing) as a
function of density, at previous epochs blue and red galaxies seem to
be mostly insensitive to the surrounding environment, with a nearly
flat distribution of the fraction of the bluest and reddest objects
over the whole over-density range.  The absence of the colour-density
relation at the highest redshift bins investigated implies that {\it
quenching of star formation was more efficient in high density
regions}.

b) {\it The evolution of the colour-density relation depends on
luminosity}.  Not only, at fixed luminosity, there is a progressive
decrease of red objects as a function of redshift in high density
regions, but also, at fixed redshift, there is a progressive decrease
of fainter red galaxies.  This result implies that {\it star formation
ends at earlier cosmic epochs for more luminous/massive galaxies}.

c) The relative fraction of the bluest objects was higher in the past
in both high and low density environments and for both more luminous
and fainter galaxies. However, the observed drop in the star formation
rate of blue objects in poor environments is weaker than in high
density environments and is weaker for fainter galaxies.  This result
indicates that star formation rate continues to be substantial at the
present day in field galaxies, and especially in the fainter ones.

d) The bimodal $(u^{*}-g')$ colour-magnitude distribution shows a marked
dependence on environment.  While locally the colour-magnitude
diagrams in low and high density regions are significantly different,
in the highest redshift bin investigated the two distributions mirror
each other. This suggests that we have sampled the relevant
time-scales over which physical {\it nurture} processes have conspired
to shape up the present-day density dependence of the colour-magnitude
distribution.

We conclude that the colour-density and colour-magnitude-density
relations are not the result of initial conditions imprinted early on
during the primordial stages of structure formation and then frozen
during subsequent evolution. Our results suggest a scenario whereby
both time evolving genetic information (galaxy biased formation) and
complex environmental interaction (star formation quenching) concurred
to build up these relations.

We remark that the scale ($5h^{-1}$Mpc) over which the over-density
field of the deep universe is reconstructed in this paper prevents a 
straightforward extrapolation of our results down to cluster scales.

In a companion paper \citep{ilbert2006} we focus on the analysis of
the environmental effects on the galaxy Luminosity Function at high
redshift. Complementary analysis based on data from various ongoing
deep redshift survey, such as the zCOSMOS \citep{lilly2006} and DEEP2
surveys \citep{Davis2003}, may help in sheding further light on our
findings. Indeed, disentangling the role of environmental conditions
on the evolution of galaxy structural parameters and especially
finding how to accommodate present observations within the context of
the competing models of galaxy formation and evolution will present a
fascinating challenge for both observers and theorists over the next
few years.

\section*{Acknowledgements}

We thank Darren Croton, Frank van den Bosch, Stefano Andreon and
Simone Weinmann for stimulating discussions. We thank also the referee
for helpful comments which improved the content of the paper. This
research has been developed within the framework of the VVDS
consortium and it has been partially supported by the CNRS-INSU and
its Programme National de Cosmologie (France), by the Italian Ministry
(MIUR) grants COFIN2000 (MM02037133) and COFIN2003 (num.2003020150)
and by PRIN-INAF 2005 (CRA 1.06.08.10). CM also acknowledges financial
support from the Region PACA. The VLT-VIMOS observations have been
carried out on guaranteed time (GTO) allocated by the European
Southern Observatory (ESO) to the VIRMOS consortium, under a
contractual agreement between the Centre National de la Recherche
Scientifique of France, heading a consortium of French and Italian
institutes, and ESO, to design, manufacture and test the VIMOS
instrument.

\bibliographystyle{aa}
\bibliography{5161bibl}

\label{lastpage}

\end{document}